# Pressure-Induced Superconductivity and Structural Transitions in Ba(Fe$_{0.9}$Ru$_{0.1}$)$_2$As$_2$

Walter Uhoya[1], Georgiy M. Tsoi[1], Yogesh K. Vohra[1], Athena S. Sefat[2], and Samuel T. Weir[3]

[1] Department of Physics, University of Alabama at Birmingham (UAB), Birmingham, AL 35294, USA

[2] Materials Science & Technology Division, Oak Ridge National Laboratory (ORNL), Oak Ridge, TN 37831, USA

[3] Mail Stop L-041, Lawrence Livermore National Laboratory (LLNL), Livermore, CA 94550, USA

**Abstract** – Electrical transport and structural characterizations of isoelectronically substituted Ba(Fe$_{0.9}$Ru$_{0.1}$)$_2$As$_2$ have been performed as a function of pressure up to ~ 30 GPa and temperature down to ~ 10 K using designer diamond anvil cell. Similar to undoped members of the $A$Fe$_2$As$_2$ ($A$ = Ca, Sr, Ba) family, Ba(Fe$_{0.9}$Ru$_{0.1}$)$_2$As$_2$ shows anomalous $a$- lattice parameter expansion with increasing pressure and a concurrent ThCr$_2$Si$_2$ type isostructural ($I4/mmm$) phase transition from tetragonal ($T$) phase to a collapsed tetragonal ($cT$) phase occurring between 12 and 17 GPa where the $a$ is maximum. Above 17 GPa, the material remains in the $cT$ phase up to 30 GPa at 200 K. The resistance measurements show evidence of pressure-induced zero resistance that may be indicative of high-temperature superconductivity for pressures above 3.9 GPa. The onset of the resistive transition temperature decreases gradually with increasing pressure before completely disappearing for pressures above ~10.6 GPa near the $T$-$cT$ transition. We have determined the crystal structure of the high-$T_c$ phase of Ru-doped BaFe$_2$As$_2$ to remain as tetragonal ($I4/mmm$) by analyzing the x-ray diffraction pattern obtained at 10 K and 9.7±0.7 GPa, as opposed to inferring the structural transition from electrical resistance measurement, as in a previous report (Ref. [6]).

**PACS: 62.50.-p, 74.62.Fj, 64.70.K-, PACS 61.50.Ks**

**Introduction**

The pressure variable has always played an important role in the discovery and optimization of properties of novel materials. Discovery of high-temperature superconductivity in a class of iron-based layered compounds under high pressure, or chemical doping, has received extensive attention [1-7]. Undoped iron-based layered compounds like $RE$OFeAs ($RE$ = rare-earth metal), and $A$Fe$_2$As$_2$ (122 type, $A$ = alkaline-earth metal) are non-superconducting at ambient pressure and are known to exhibit tetragonal to orthorhombic structural transition and antiferromagnetic (AFM) ordering on cooling. Both the structural transition and AFM ordering in 122 material can be suppressed under pressure or chemical substitution with the appearance of superconductivity occurring at low temperatures [1-14].

The 122 Fe-based materials have ThCr$_2$Si$_2$-type $T$ crystal structure at ambient conditions, and undergo a pressure-induced structural transition to a $cT$ phase, originally reported for CaFe$_2$As$_2$ [15-17]. The $T$-$cT$ phase transition has been reported in other 122 materials at ambient temperature and is accompanied by anomalous $a$-lattice parameter expansion concurrently with a collapse of $c$-lattice parameter [11, 18 - 22].

Structural, magnetic, and electronic properties for CaFe$_2$As$_2$ member of 122 materials are well studied; revealing relationships among the antiferromagnetic, superconducting, and structural phases in the pressure-temperature (P-T) phase diagram as discussed in recent review [23]. At ambient pressure, CaFe$_2$As$_2$ undergoes a first order phase transition from a high temperature $T$ phase to a low-temperature orthorhombic/antiferromagnetic phase upon cooling through T ~170 K. With the application of pressure, this phase transition is rapidly suppressed and by ~0.35 GPa, it is replaced by a first order phase transition to a low-temperature $cT$, non-magnetic phase. Further application of pressure leads to an increase of the $T$ to $cT$ phase transition temperature, with it crossing room temperature by ~1.7 GPa [23]. Early studies have suggested that the appearance of $cT$ phase in CaFe$_2$As$_2$ is closely related to the onset of pressure-induced superconductivity [15 – 17]. However, the $cT$ phase in these studies was found under non-hydrostatic pressure generated by a liquid medium that solidified on cooling [15]. This non-hydrostaticity leads to artificial broadening of the $cT$ phase, multi-phase behavior and a superconducting phase that span the critical pressure where the collapse occurs [23, 24]. In contrast, experiments under highly hydrostatic pressure conditions (He-gas pressure transmitting medium) on pure CaFe$_2$As$_2$ have revealed that neither the ambient pressure orthorhombic phase nor the *cT* phase support superconductivity [25]. These results further suggest that superconductivity in CaFe$_2$As$_2$ may be associated with a low temperature multi-crystallographic phased sample resultant of non-hydrostatic conditions associated with the combination of a first-order structural phase transition and frozen liquid media [25]. Recently, simultaneous resistivity measurements and neutron diffraction studies under uniaxial pressure have strongly suggested that the high-temperature *T* phase is stabilized by uniaxial stress down to low temperatures, and this tetragonal phase may be responsible for the superconductivity [24]. High-pressure studies for other 122 parent materials have shown that pressure-induced superconductivity occurs in the *T* phase and vanishes at similar pressures that *cT* phase appears [18 - 21]. Apart from the application of high pressure, *T* to *cT* phase transition has been induced through chemical pressure [3, 26]. Saha et al. reported that chemical substitution of Ca$^{2+}$ with *RE* ion substituent is enough to cause crystal structure of Ca$_{1-x}$*RE*$_x$Fe$_2$As$_2$ to collapse when the interlayer As-As anion separation falls below 3 Å [26]. Superconductivity has also been induced in 122 materials by the chemical substitution of *A* or transition metal [3, 6, 8 -10], resulting in $T_c$ as high as ~45 K [3, 26]. Recent studies on lightly doped Ba(Fe$_{2-x}$Ru$_x$)$_2$As$_2$, x < 0.15 [6, 12-14] samples have shown no evidence of superconductivity at ambient pressure. However, the partial chemical substitution of Fe with Ru in Ba(Fe$_{2-x}$Ru$_x$)$_2$As$_2$ with x > 0.15 [6, 12-14] leads to the suppression of tetragonal to othorhombic stuctural transition, and superconductivity is induced with optimal $T_c$ of ~ 17 K for range of x ≈0.25 to 0.3 at ambient pressure. Resistivity data on Ba(Fe$_{2-x}$Ru$_x$)$_2$As$_2$ for x < 0.15 samples have been presented for pressures up to 2 GPa using a piston cylinder cell (PCC) with a 4:6 = light mineral oil:n-pentane, and for pressures up to 5 GPa using modified Bridgman cylinder cell (BC) with a 1:1 mixture of isopentane:n-pentane liquid pressure transmitting media [6]. The high-pressure part of the phase-diagram of these samples could not be investigated because of pressure limits of the two pressure cells and low hydrostatic limit of the liquid pressure transmitting media, which solidifies at pressures less than 6.5 GPa and ambient temperature [6]. We have overcome the above limitations by determining the higher-pressure part of the phase diagram of Ba(Fe$_{2-x}$Ru$_x$)$_2$As$_2$ using designer diamond anvil cell (DAC), which can be used for transport and structural property investigations up to much higher pressures.

Additionally, the effects of chemical doping and high pressure on transport and structural properties have been studied for Ba(Fe$_{2-x}$Ru$_x$)$_2$As$_2$ up to 4.9 GPa, and it was shown that 3 GPa is equivalent to 10% Ru substituent in Fe site [6]. However, the structural transition reported in these studies were derived from temperature-dependence of resistance alone under the assumption that the AFM ordering and structural transition, known to be concurrent for 122 materials at ambient pressure, remains unseparated under pressure. However, this is not always true, as it was shown that the two transitions can be decoupled under high pressure or chemical doping for some 122 materials [7]. Systematic neutron or x-ray diffraction experiments at low temperatures would be necessary to clearly determine the crystallographic structures of these materials at high pressure.

At present, high pressure crystallographic data for any Ru-subtituent concentration x, in $Ba(Fe_{2-x}Ru_x)_2As_2$; $0 < x < 1$, have not been been presented and this is needed to correlate crystallographic properties with superconducting and magnetic properties observed under high pressures. In particular, pressure-induced $T_c$ with a maximum of ~ 25.7 K at 4.94 GPa has been reported for lightly-doped $Ba(Fe_{2-x}Ru_x)_2As_2$ with $x < 0.15$ [6]. $T_c$ seems to be maximum around 4.94 GPa pressure limit reported for x=0.09 [6]. Hence electrical transport measurements on these materials to pressures beyond 5 GPa would be necessary to completely determine the higher pressure part of the superconductivity phase diagram. Rather than inferring the structual transition of the high pressure superconducting phase of these materials [6], here we have analyzed the structure using x-ray powder diffraction of the high pressure superconducting phase of $Ba(Fe_{0.9}Ru_{0.1})_2As_2$. Although the occurrence of negative lattice parameter compressibility phenomenon associated with *cT* transitions in compounds with 122 $ThCr_2Si_2$-type structure has not been systematically reported at low temperatures, pressure-induced changes in superconducting and magnetic properties of Fe-based materials [7, 11] suggest that unit-cell parameter changes may have a direct bearing on superconducting and magnetic properties of these systems. Therefore, the pressure-dependence of unit cell parameters for Ru-doped $BaFe_2As_2$ has been investigated at low temperatures.

In this work, we have investigated the high-pressure effects on the crystal structure of the layered $ThCr_2Si_2$-type $Ba(Fe_{0.9}Ru_{0.1})_2As_2$ up to 30 GPa and low temperatures down to 10 K using synchrotron x-ray diffraction technique in a DAC. These studies were complemented by pressure and temperature-dependent electrical resistance measurements on the same sample using a designer DAC up to 17 GPa and down to 10 K. The x-ray diffraction data collected reveal a highly anisotropic and anomalous compressibility effects in which *a*-axis increases with increasing pressures up to a maximum then decreases normally whereas the *c*-axis decreases continuously with increasing pressure. Analysis of the x-ray diffraction data obtained for Ba(Fe$_{0.9}$Ru$_{0.1}$)$_2$As$_2$ at 200 K indicates a pressure induced $T$ to $cT$ phase transition. This transition occurs between ~12 GPa that marks the highest measured pressure where the sample is in $T$ phase and ~ 17 GPa, which marks the lowest measured pressure where the sample is in the $cT$ phase, and the crystal structure remains in the $cT$ phase up to 30 GPa. The resistance measurements under high pressure show existence of an onset of resistive anomaly upon lowering temperature, and zero resistance at lower temperatures. The resistive anomaly is reminiscent of pressure induced superconducting resistive transition reported for 122 materials. Evolution of $T_{c\ zero}$ and $T_{c\ onset}$ determined in the present work, using solid steatite as a pressure transmitting medium, is consistent with previous studies for a nearly similar Ru doped BaFe$_2$As$_2$ under liquid pressure medium [6], but the present $T_c$ values are shifted to lower temperatures. The onset of $T_c$ decreases gradually with increasing pressure before completely dissappearing for pressures above 10.6 ± 0.7 GPa. The loss of superconductivity nearly coincides with the critical pressure where the sample transforms from $T$ to $cT$ structure reported in x-ray diffraction studies at 200 K, and this is consistent with the early works that suggest $cT$ phase in non-superconducting CaFe$_2$As$_2$ [24, 25] and Ba, Sr analogues [18-22]. Analysis of the x-ray diffraction data under high pressure and low temperature conditions suggest that the crystal structure of the superconducting phase remain tetragonal ($I4/mmm$) of ThCr$_2$Si$_2$ type. The present study provides further experimental evidence that pressure can be used to induce superconductivity in lightly doped Ba(Fe$_{2-x}$Ru$_x$)$_2$As$_2$, x < 0.15, which are otherwise non-superconducting at ambient pressure.

**Experimental Details**

Large platelets of Ba(Fe$_{02-x}$Ru$_x$)$_2$As$_2$ (x=0.1±0.05) single crystals were grown out of a mixture of Ba, FeAs flux, and RuAs similar to that described in reference [8], with crystallographic [001] direction perpendicular to their crystalline plate surface. The crystals were characterized using energy dispersive x-ray spectroscopy (EDS), x-ray diffraction (XRD), temperature-dependent magnetization and electrical resistance measurements. Elemental analysis of single-crystal samples was used to determine the actual percentage of the Ru-substituent level (x) in the lattice. Chemical compositions were found by averaging EDS data taken from three different spots on the surface of each single crystal. The estimated error on each EDS x values is ~ 5 %. At room temperature, the structures were identified as the tetragonal ThCr$_2$Si$_2$ structure type (I4/mmm). The physical properties at ambient pressure showed an anomaly corresponding to antiferromagnetic transition at $T_N$ = 95 K, with no evidence of superconducting transition down to 2 K. Electrical resistance measurements at high pressures were performed using four-probe method in an eight tungsten microprobe designer diamond anvil cell as described earlier [27, 28]. The eight tungsten microprobes are encapsulated in a homoepitaxial diamond film and are exposed only near the tip of the diamond to make contact with the sample at high pressure. The sample was loaded into a 120 μm hole of a spring-steel gasket that was first pre-indented to a ~95-μm thickness and mounted between a matched pair of the diamond anvils ready for high-pressure x-ray diffraction experiments. Two electrical leads pass constant current through the sample and two additional leads measure voltage across the sample. Care was taken to electrically insulate the sample and the designer microprobes from the metallic gasket by using solid steatite medium. In addition, the solid steatite acts as a pressure transmitting medium, so the present studies can be regarded as non-hydrostatic, where the small uniaxial stress component is limited by the sample shear strength at high pressures. Even though solid pressure medium provide less homogenous pressure, it is advantagious for our mega-bar experiments in DAC, unlike liquid medium which solidify at higher pressures changing hydrostaticity of sample condition drastically. For instance, the hydrostatic limit for the pressure medium used in the previous high pressure work on $Ba(Fe_{1-x}Ru_x)_2As_2$ single crystal was only 6.5 GPa [6]. In our experiments, pressure was applied using a gas membrane to the designer DAC and monitored using ruby fluorescence techniques as described in our earlier publication [28]. For simultaneous temperature- and pressure-dependent x-ray diffraction experiments, the designer DAC was cooled in a continuous helium flow-type cryostat, and the pressure in the cell was measured *in situ* with the ruby fluorescence technique [28-29]. The synchrotron XRD experiments were performed at the high pressure beam-line 16-BM-D of HPCAT, at the Advanced Photon Source in Argonne National Laboratory. An angle dispersive technique with a MAR345 image-plate area detector was employed using a focused monochromatic beam with x-ray wavelength, λ = 0.424602 Å and sample to detector distance of 313.1 mm. The image plate XRD patterns were recorded with a focused x-ray beam of 6 μm by 13 μm (fwhm) on an 80 μm diameter sample mixed with ruby to serve as pressure marker. Experimental geometric constraints and the sample-to-image plate detector distance were calibrated using $CeO_2$ diffraction pattern and were held at the standard throughout the entirety of the experiment. The software package FIT2D [30] was used to integrate the collected MAR345 image plate diffraction patterns which were analyzed by GSAS [31] software package with EXPGUI interface [32] employing full-pattern Rietveld refinements and Le Bail fit techniques to extract structural parameters.

**Results and Discussions**

Figure 1 shows representative x-ray diffraction patterns of Ba(Fe$_{0.9}$Ru$_{0.1}$)$_2$As$_2$ obtained at various temperatures and pressures. Figure 1 (a) shows Rietveld refinement of the x-ray diffraction pattern at ambient temperature and 1.7 GPa, revealing the tetragonal *I*4/*mmm* with lattice parameters $a = b = 3.9644(5)$ Å, $c = 12.7594(3)$ Å, $v = 200.5$ Å$^3$, As-As distance of 3.7002 Å and an axial ratio ($c/a$) = 3.2185(8). The unit cell has Ba atoms at the 2*a* position (0, 0, 0), Fe/Ru atoms at the 4*d* positions (0, 1/2, 1/4) and (1/2, 0, 1/4), and As atoms at the 4*e* positions (0, 0, $z$) and (0, 0, -$z$), with refined value of $z = 0.355$ at 1.7 GPa. The diffraction lines are all identified and indexed and the difference between the observed x-ray diffraction pattern and Rietveld fit is satisfactorily small suggesting a single-phase sample with ThCr$_2$Si$_2$-type crystal structure, same as the ambient pressure 122 structure [4-7]. Figure 1 (b) shows an x-ray diffraction pattern of Ba(Fe$_{0.9}$Ru$_{0.1}$)$_2$As$_2$ at 10 K and 9.7 GPa where the sample is superconducting, as discussed later in this report. The temperature-dependent magnetization and resistivity measurements on Ba(Fe$_{0.9}$Ru$_{0.1}$)$_2$As$_2$ at ambient pressure have shown an anomaly corresponding to antiferromagnetic and structural (tetragonal to orthorhombic) transitions at 95 K, with no evidence of superconducting transition down to 2 K. The antiferromagnetic/structural transition temperature is slightly less than that for Ba(Fe$_{0.91}$Ru$_{0.09}$)$_2$As$_2$ (~98 K, Ref. [6]) as compared to that for undoped BaFe$_2$As$_2$ with the transition at 134 K [6]. As expected, both compounds show no evidence of superconductivity at ambient pressure. At 10 K and 9.7 GPa, the orthorhombic peaks expected below 95 K are absent and the low temperature x-ray diffraction pattern is similar to the room temperature tetragonal pattern (figures 1 (a-b)). The Rietveld refinement of the XRD pattern confirmed a structure with $a = b = 4.0286(4)$ Å, $c = 11.2108(10)$ Å, $v = 181.95(8)$ Å$^3$ and refined $z = 0.358$, and As-As = 3.1839 Å. The observation of the ThCr$_2$Si$_2$-type crystal structure at 10 K suggests a complete suppression of the structural transition and AFM ordering and that the superconducting phase is predominantly tetragonal. The As-As bond distance of 3.1839 Å obtained at 10 K and 9.7 GPa is slightly larger than the critical value of ~ 3 Å which has been determined to control the *T* to *cT* phase transition in 122 materials [26].

When comparing the room temperature diffraction pattern at 1.7 GPa (figure 1(a)) with the pattern at 9.7 GPa and 10 K (figure 1(b)), the Bragg (hkl) peaks that depend only on the length of lattice parameter *a,* such as (110), (200) and (220) are strongly shifted to lower diffraction angles or higher d-spacing. In contrast, the (hkl) peaks that are strongly dependent on the *c*-axis length such as (103) and (112) shift to higher diffraction angles and the peaks that dependent on both *a*- and *c*- lattice parameters remain approximately in the same position. Conventional behavior of material under compression suggests that all peaks would move to higher angles in the x-ray diffraction spectrum; however, in comparing figure 1(a) with 1(b), all *a*-dependent peaks seem to move to lower diffraction angles for higher pressures. This is a clear indication of negative compressibility in the *a*-axis of the tetragonal lattice structures for Ba(Fe$_{0.9}$Ru$_{0.1}$)$_2$As$_2$. Expansion of the tetragonal lattice parameter *a*- upon compression has been reported for a number of 122 materials under high pressure and ambient temperature [11, 16, 19-23, 26] and this effect occurs prior to structural transition to the collapsed tetragonal phase [16]. A controlled experiment at constant temperature is desirable to investigate pressure-induced changes on unit cell parameters, hence we have performed further x-ray diffraction experiments at 200 K and various pressures up to 30 GPa.

Figure 2 (a-d) shows measured tetragonal unit cell parameters *a* and *c*, axial ratio *c/a* and unit cell volume (*v*) that were obtained for Ba(Fe$_{0.9}$Ru$_{0.1}$)$_2$As$_2$ under high pressures up to 30 GPa at 200 K. The error bars in pressure were determined from the full width at half maximum of ruby R-1 spectral line used to determine pressure as discussed previously. Upon increasing pressure, anomalous compression effects are observed in which the *a*-lattice parameter expands rapidly up to a maximum between ~12 and ~17 GPa. Above 17 GPa, *a*-lattice parameter decreases normally with increase in pressure up to ~ 30 GPa.

The *c* lattice parameter however shows a rapid decrease with increasing pressure in the same pressure range where *a*-axis increases as shown in figure 2 (b). The pressure evolution of the *c*-axis changes from rapid below 12 GPa to gradual above 17 GPa, suggesting a break in the slope of the *c*-lattice parameter between the two pressure values. This break point suggests a collapse in the tetragonal lattice parameter *c* -, and so it defines the pressure-induced phase transition from *T* to the *cT* phase as discussed in the next section. The measured unit cell parameters at 11.77 GPa, below which the *c* decrease more rapidly with increase in pressure are *a* = 4.0297 Å, *c* = 10.8661 Å, *c/a* = 2.6965 and *v* = 176.4490 Å$^3$; at 16.79 GPa, beyond which the sample is in the *cT* phase and exhibit a gradual decrease in *c* are *a* =4.0303 Å, *c* =10.3424 Å, *c/a* = 2.5661 and *v* =167.9948 Å$^3$. It can be seen from these measurements that the unit cell volume decreases by ∼ 5 % across the transition (region between dotted line in figure 2 (b)). The x-ray diffraction measurements for CaFe$_2$As$_2$ have shown [16, 17] that during the structural phase transition from the tetragonal to the collapsed phase, the *c*-lattice parameter contracts by about ∼9%.

The As-As interlayer separation was determined to be the key parameter controlling the isostructural *T-cT* transition for 122 materials: CaFe$_2$As$_2$ collapses once the interlayer As-As distance reaches a critical value of ∼ 3 Å where As *p*-orbitals overlap upon application high pressure or rare-earth doping [26]. The measured As-As distance for our sample at 11.77 GPa and 200 K is 3.0860 Å confirming that the compressed sample is in the tetragonal structure phase below 12 GPa with As-As > 3 Å. However, the measured As-As distance for our sample at 16.79 GPa and 200 K is 2.9372Å, which is quite less than 3 Å and thus suggests that the high pressure crystal structure phase for pressures above 17 GPa is indeed in the collapsed tetragonal phase. Because of difficulty in controlling the membrane pressure for DAC at 200 K, we could not collect data between 12 and 17 GPa and hence could not accurately determine the transition pressure. We therefore use the break in the slope of *c*-axis to correspond to the critical value for the *T-cT* phase transition pressure. This transition occurs at ∼14 GPa as determined by the intersection of the two linear fits for the *c* vs. pressure curves as shown in figure 2 (b). The estimated *c*-lattice parameter at 14 GPa (figure 2 (b)) is 10.4762 Å. Since no x-ray diffraction patterns were measured at the transition pressure, we estimated As-As bond distance at 14 GPa to be 2.9752 Å using the relation As-As = (1-2z)c, where *c* = 10.4762Å. The z-positional parameter at the transition pressure is taken to be 0.358 since refined z does not change considerably between 1.7 GPa (z=0.355) and 10 GPa (z=0.358).

Figures 2 (c) and 2 (d) show the *c/a* ratio and the unit cell volume (*v*) as a function of pressure in *T* and *cT* phases. For each figure, the dotted line at ∼12 GPa corresponds to maximum measured pressure for *T*-phase, while that at ∼17 GPa corresponds to minimum measured pressure for *cT* phase. While no clear discontinuity is visible in the volume vs. pressure curve, a clear change is observed between 12 and 17 GPa for *c/a* vs pressure curve, similar to that on *c*-lattice parameter vs. pressure curve. The *T-cT* transition pressure is determined to be ~ 14 GPa by the criterion described above for the *c*-axis and illustrated in figure 2 (b). Recent studies on the effect of pressure on 122 materials by joint experimental and theoretical computation have shown that *a*-axis expansion and the corresponding collapse in *c* and *c/a* can be explained in terms of stabilization of the As-As bonding states that stabilizes Fe-As antibonding across *T-cT* phase transition [20]. Applying pressure leads to a decrease of the distance between FeAs layers along the *c*-axis and therefore As-As separation. At sufficient pressure, critical As-As distance is reached enabling a sufficiently high enough orbital overlap of the 4$p_z$ orbitals, promptly the bonding interactions dominate and the tetragonal to collapsed phase transition occurs. The As-As distance corresponding to *T-cT* phase transition has been determined to be 3 Å [26]. We have shown that for Ba(Fe$_{0.9}$Ru$_{0.1}$)$_2$As$_2$, the *T-cT* phase transition occurs at ~14 GPa with corresponding As-As bond distance of 2.9752 Å. Electronic structure calculations further outline a scenario where a stabilization of the As-As bonding states stabilizes Fe-As antibonding states which have lifted up the former ones to energies near the Fermi level. Consequently, the Fe-As bonds become weaker and the lattice parameter *a* increases whereas *c* decreases across the phase transition [20].

Figure 3 shows temperature dependence of the electrical resistance of Ba(Fe$_{0.9}$Ru$_{0.1}$)$_2$As$_2$ at various pressures. At low temperatures, the resistance measurements show a downturn resistive anomaly labelled as $T_{c\ onset}$ (figure 3), and this anomaly persists up to nearly 8.5 GPa with no evidence of bulk superconductivity. The downturn resistive anomaly is similar to that observed in parent materials BaFe$_2$As$_2$ and SrFe$_2$As$_2$ and was associated with some form of strain-induced superconductivity in a very small fraction of the materials [5]. While no evidence of superconducting resistive transition with zero resistance was found from the electrical resistance measurements at pressures above 6 GPa, a resistive drop to zero resistance is clearly observed at low pressures between 3 and 5 GPa. The observed pressure induced zero resistance state could be an indication of high-$T_c$ superconductivity given that similar resistive anomaly has been associated with superconductivity in *A*Fe$_2$As$_2$ materials [7, 11]. The inset in figure 3 illustrates the criteria used for defining the onset of the resistive transition ($T_{c\ onset}$) and the zero resistance temperature ($T_{c\ zero}$) for this material. The $T_{c\ onset}$ for ~3.9±0.3 GPa occurs at 23±1 K with zero resistance occurring at $T_{c\ zero}$ = 14.5 K. On increasing pressure, the $T_{c\ onset}$ gradually shifts to lower temperatures and eventually vanishes for pressures above 10.6±0.7 GPa (figure 3). The $T_{c\,zero}$ is only observed on a narrow pressure range (3 GPa < P < 5 GPa), consistent with Ref. [6]. Moreover, the measured resistance never reaches zero for P > 6.1±0.4 GPa, and the transition width of the $T_{c\,onset}$ broadens considerably with increasing pressure; and $T_{c\,onset}$ seems to be suppressed between 8.5 and 10.6 GPa. Further increase in pressure leads to a gradual increase in the electrical resistance of the sample at low temperatures without any evidence of superconducting resistive behavior up to 17.4 GPa. Since the present measurements are restricted to temperatures above 10 K, it is possible that the $T_{c\,onset}$ and zero resistance occur below 10 K for higher pressures. The fact that superconducting transition signal broadens considerably between 8.5 GPa and 10.6 GPa could be indicative of a possible phase transformation to a non-superconducting $cT$ phase as discussed later in the text. Additional effect of non-hydrostatic pressure gradient may cause structural phase mixture hence additional broadening of resistive superconducting transition temperatures. In fact experiments under uniaxial pressure have strongly suggested that the high-temperature $T$ phase for $CaFe_2As_2$ is stabilized by non-hydrostatic component of pressure down to low temperatures [24] and this non-collapsed tetragonal phase is responsible for the superconductivity observed in the 122 material under non-hydrostatic pressure conditions [23, 24]. Furthermore, systematic high-pressure experiments under highly hydrostatic conditions have not revealed any evidence of superconductivity in either the orthorhombic phase or $cT$ phase [25].

Figure 4 shows a clearer depiction of the pressure evolution of the superconducting transition temperatures, $T_{c\,zero}$ and $T_{c\,onset}$. The vertical error bars for $T_{c\,onset}$ (figure 4) correspond to half the transition width (W) estimated at points where electrical resistance value is 25% lower than that determined at the measured onset for the superconducting transition as shown in the inset in figure 3. The $T_{c\,zero}$ increases slightly with increasing pressure from 15 K at 3.9 GPa to 17 K at 4.2 GPa, and completely vanishes for pressures greater than 6.1 GPa. The maximum $T_{c\,zero}$ (17 K) from the present work is consistent with that at ambient pressure for $Ba(Fe_{1-x}Ru_x)_2As_2$ obtained for x = 0.29 (16.5 K) [14]. Also plotted in figure 4 are Kim et al data for $T_{c\,zero}$ and $T_{c\,onset}$ for $Ba(Fe_{0.91}Ru_{0.09})_2As_2$ obtained using liquid pressure transmitting media up to 5 GPa [6]. The dependence of $T_{c\,zero}$ on pressure is qualitatively similar to that for $Ba(Fe_{0.91}Ru_{0.09})_2As_2$ [6] under liquid medium, in which pressure induced $T_{c\,zero}$ occurs at 3.16 GPa, and was found to be restricted on a narrow pressure range of 3.16 to 4.94 GPa.

The $T_{c\ onset}$ remains constant between 3.9 and 6.1 GPa, and then decreases gradually with further increasing pressure up to 10.6 GPa above which it could not be detected. The measured $T_{c\ onset}$ reported by Kim et al for Ba(Fe$_{0.91}$Ru$_{0.09}$)$_2$As$_2$ [6] using liquid pressure medium is included for comparison. A simple extrapolation of a quadratic fit to data (figure 4) suggests that the $T_{c\ onset}$ is suppressed at ~11 GPa, which is consistent with the measured 10.6±0.7 GPa where the observed $T_{c\ onset}$ is suppressed (see figure 3). This pressure is near the $T$-$cT$ phase transition pressure for Ba(Fe$_{0.9}$Ru$_{0.1}$)$_2$As$_2$, which we have determined to occur between 12 and 17 GPa. Furthermore, analysis of the XRD pattern obtained from the sample in the superconducting phase at 10 K and 9.7 ± 0.7 GPa has revealed a non-collapsed tetragonal crystal structure with As-As bond distance of 3.1839 Å. This distance is slightly larger than the critical value of ~ 3 Å where the $T$-$cT$ structural transition is expected to occur. The present results suggest that superconductivity observed in Ru doped BaFe$_2$As$_2$ under non-hydrostatic pressure conditions occurs in the non-collapsed tetragonal phase (same structure as room temperature ThCr$_2$Si$_2$-type) that is stabilized to lower temperatures by the non-hydrostatic pressure generated in the DAC. The occurrence of superconductivity in the non-collapsed tetragonal phase of Ru-doped BaFe$_2$As$_2$ is consistent with previous reports on analogous materials, $A$Fe$_2$As$_2$ ($A$=Ba, Sr) [7, 11, 18-21] and CaFe$_2$As$_2$ [23-25]

A comparison of our results with the high pressure results for Ba(Fe$_{0.91}$Ru$_{0.09}$)$_2$As$_2$ (figure 4) shows some remarkable differences and similarities. First, the variation of $T_{c\ onset}$ under high pressure for Ba(Fe$_{0.9}$Ru$_{0.1}$)$_2$As$_2$ is qualitatively similar to that for Ba(Fe$_{0.91}$Ru$_{0.09}$)$_2$As$_2$, which also depicts a constant $T_{c\ onset}$ values for pressures ranging from 2.58 to 4.94 GPa. In both cases, $T_c$ rises slightly with increase in pressure to a maximum and the $T_c$ vs pressure curves seem to form half-dome shape. The $T_c$-dome for the data obtained from solid media seems to be slightly shifted to the right as compared to that for data obtained using liquid medium. Unfortunately the measurements from Ref [6] do not go to high enough pressures to suppress the zero-resistivity transition and onset of the resistive transition. Additionally, the $T_c$ values for Ba(Fe$_{0.9}$Ru$_{0.1}$)$_2$As$_2$ are significantly lower in comparison to those for Ba(Fe$_{0.91}$Ru$_{0.09}$)$_2$As$_2$ (see figure 4). Since the concentration of the substituent in the two samples are nearly equal, the observed differences could be likely attributed to pressure inhomogeneity because the two measurements were performed under different hydrostatic conditions: Ba(Fe$_{0.9}$Ru$_{0.01}$)$_2$As$_2$ in DAC with solid pressure medium (present study) as compared to BC with liquid medium for Ba(Fe$_{0.91}$Ru$_{0.09}$)$_2$As$_2$ [ 6]. The pressure inhomogeneity for the sample in DAC (present work) was estimated from ruby pressure marker and is plotted as horizontal error bar in in $T_c$ onset vs pressure curve (figure 4). The horizontal error bars in pressures were determined from the full width at half maximum (fwhm) of Ruby fluorescence R-1 line used to determine pressure, and suggest that the pressure inhomogeneity inside the designer DAC increases with pressure. At higher applied pressures, the degree of pressure non-hydrostaticity is quite high and possibly results in the broadening of the width of the superconducting transition as discussed earlier (figure 3). Since the high pressure experiments in our studies were carried out using solid pressure medium with the *c*-axis of the sample normal to the anvil load, the uniaxial pressure component is expected to be large as compared to the setup from Ref [6] where liquid medium was used with the PCC and BC up to 5 GPa. The increased uniaxial pressure component may explain the loss of zero resistance at ~ 6 GPa and the appearance of a region of broad-non bulk superconductivity with broad resistive transition for higher pressures.

The pressure induce phase transitions in 122 materials have been shown to be quite sensitive to the degree of non-hydrostaticity of the pressure transmitting media (see detail reviews in Refs. [7, 11, 33]. The evolution of the electrical resistivity of $BaFe_2As_2$ single crystals with pressure up to 5 GPa have been investigated using three different pressure-transmitting media, namely pentane-isopentane, daphne oil and steatite with different intrinsic level of hydrostaticity and yield moderate inhomogeneous pressure distributions at very low pressure [33]. These studies showed that that the pressure-temperature phase diagram of $BaFe_2As_2$ is extremely sensitive to the pressure-transmitting medium used for the experiment and, in particular, to the level of resulting uniaxial stress. An increasing uniaxial pressure component in this system quickly reduces the spin-density-wave AFM order and favors the appearance of superconductivity. In addition to non-hydrostatic stress and strain induced differences, further difference can be attributed to the Ru-dopant disordering since the Ru-dopant level for the two compounds is slightly different. Furthermore the resistive anomalies in our present work are rather broad in applied pressure compared to that in Ref. [6] and different criteria for the definition of onset transition temperatures can shift temperature assignments.

**Conclusions**

In conclusion, we have presented data and analysis of synchrotron x-ray diffraction measurements on a single crystalline Ba(Fe$_{0.9}$Ru$_{0.1}$)$_2$As$_2$ as a function of applied pressures up to 30 GPa and temperatures down to 10 K. The XRD studies were complemented with a series of temperature and pressure dependence of electrical resistance on the same sample, for pressures up to ~ 17 GPa and temperatures down to ~10 K. The studies were performed in DAC using solid pressure pressure transmitting. We found that Ba(Fe$_{0.9}$Ru$_{0.1}$)$_2$As$_2$ shows a negative *a*-lattice parameter compressibility and *T-cT* phase transition occurring between 12 and 17 GPa where the *a*-lattice maximizes. The resistance measurements for pressures above 3 GPa show a broad region of non-bulk or partial superconductivity, which is characterized by broad electrical resistive transition that does not even become zero for P = 6 GPa up to 8.5 GPa. However, zero resistance exists within a narrow pressure range (between 3 and 5 GPa). Comparison of present work from solid medium and previous work from liquid medium suggests that the evolution of $T_{c\ zero}$ and $T_{c\ onset}$ with pressure is strongly affected by the hydrostaticity of the pressure-transmitting medium. The measured onset of the superconducting transition is suppressed between 8.5 ± 0.6 and 10.6±0.7 GPa. The loss of superconductivity nearly coincides with the critical pressure where the sample transform from *T* to *cT* structure, reported in x-ray diffraction studies at 200 K. The crystal structure of the high-$T_c$ phase of Ru-doped BaFe$_2$As$_2$ is tetragonal of ThCr$_2$Si$_2$ type, found by analyzing the x-ray diffraction pattern obtained at 10 K and 9.7±0.7 GPa; *cT* phase is obtained at higher pressures.

**Acknowledgment**

Walter Uhoya acknowledges support from the Carnegie/Department of Energy (DOE) Alliance Center (CDAC) under grant no. DE-NA0002006. Part of this work is based upon work supported by the Department of Energy National Nuclear Security Administration under Award Number DE-NA0002014. Portions of this work were performed in a synchrotron facility at HPCAT (Sector 16), Advanced Photon Source (APS), Argonne National Laboratory. The work at ORNL was supported by the Department of Energy, Basic Energy Sciences, Materials Sciences and Engineering Division.

**Figure Captions**

**Figure 1.** (Color online) Rietveld refinement of powder x-ray diffraction patterns of Ba(Fe$_{0.9}$Ru$_{0.1}$)$_2$As$_2$ in the tetragonal phase at (a) ambient temperature and 1.7 GPa, and at (b) 10 K and 9.7 GPa. The lower most solid line (magenta) in (a) and (b) are the difference profile curves between the observed (solid red symbols) and calculated (green line) profiles. The (hkl) values for peaks corresponding to the tetragonal *I*4/*mmm* phase are marked.

**Figure 2.** (Color online) Measured tetragonal *a* and *c* lattice parameters, the axial ratio *c/a* and unit cell volume (*v*) for Ba(Fe$_{0.9}$Ru$_{0.1}$)$_2$As$_2$ as a function of applied pressure. The x-ray diffraction measurements at high pressures were performed at a low temperature of 200 K. The solid curves in (b) are linear fits to *c* for the *T* phase at low pressures and *cT* at higher pressures. The vertical lines at 12 GPa and 17 GPa in each graph mark the highest measured pressure below which the sample is in the *T* phase and the lowest measured pressure above which the sample is in the *cT* phase respectively.

**Figure 3.** (Color online). Temperature dependence of the electrical resistance of Ba(Fe$_{0.9}$Ru$_{0.1}$)$_2$As$_2$ at various applied pressures. Steatite was used a pressure medium. Criteria used to determine the onset of superconducting transition temperature $T_{c\ onset}$, the zero resistance $T_{c\ zero}$, and the transition width (W) are illustrated in the inset figure for P = 3.9 GPa.

**Figure 4.** (Color online). Measured superconducting transition temperature for Ba(Fe$_{0.9}$Ru$_{0.1}$)$_2$As$_2$ as a function of pressure, by onset temperature $T_{c\ onset}$ criterion. The solid curve is a quadratic fit to the data and is described in the text. Vertical error bars are determined from the transition width and horizontal error bars are determined by the FWHM of the ruby fluorescence line as described in the text. Results for Ba(Fe$_{0.91}$Ru$_{0.09}$)$_2$As$_2$ from Ref. [6] are included for comparison.

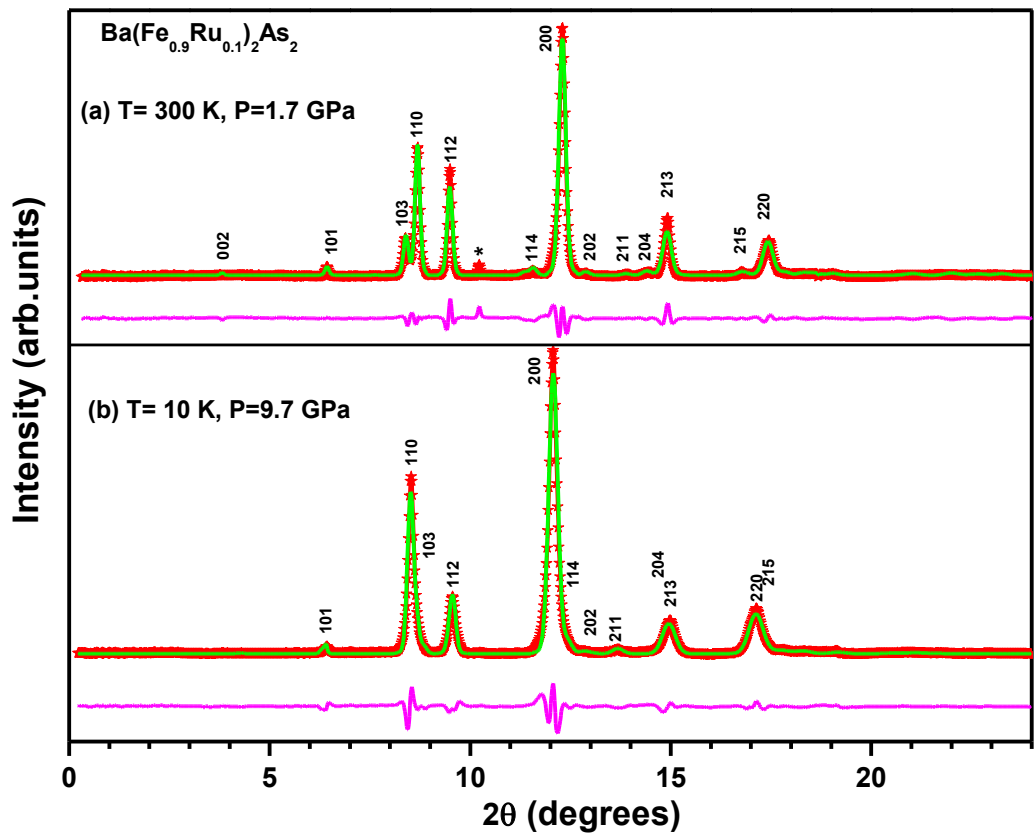

Figure 1

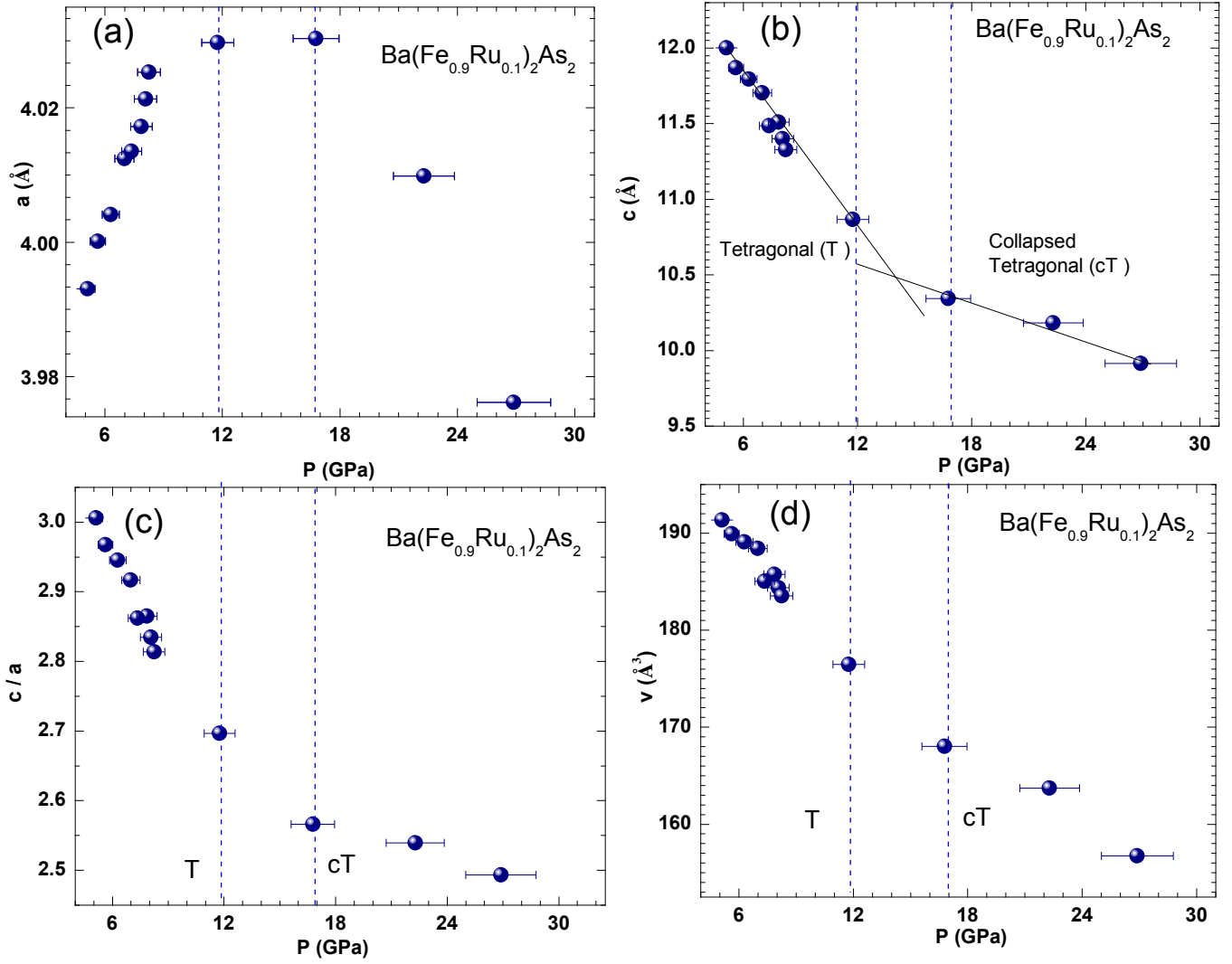

Figure 2

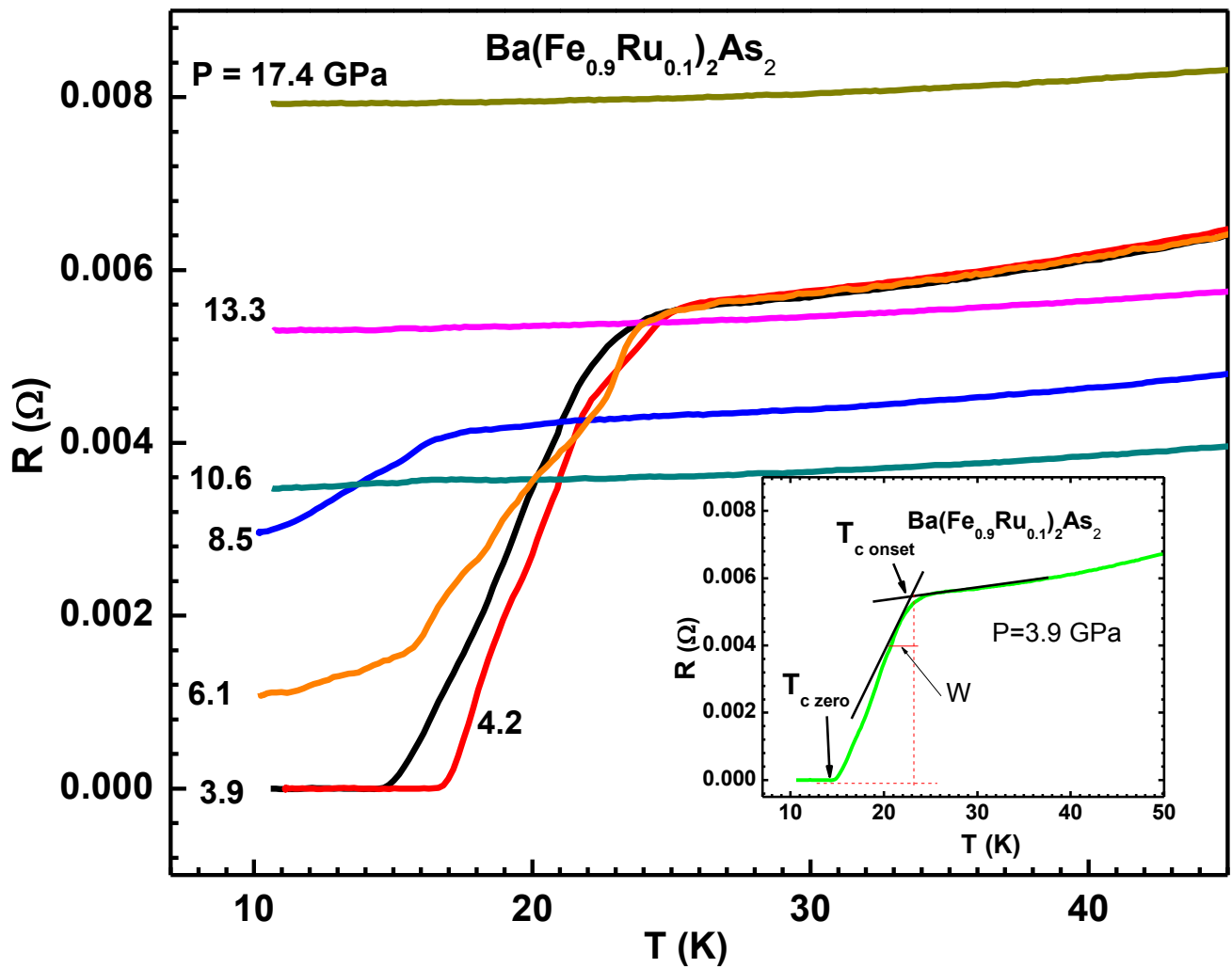

Figure 3

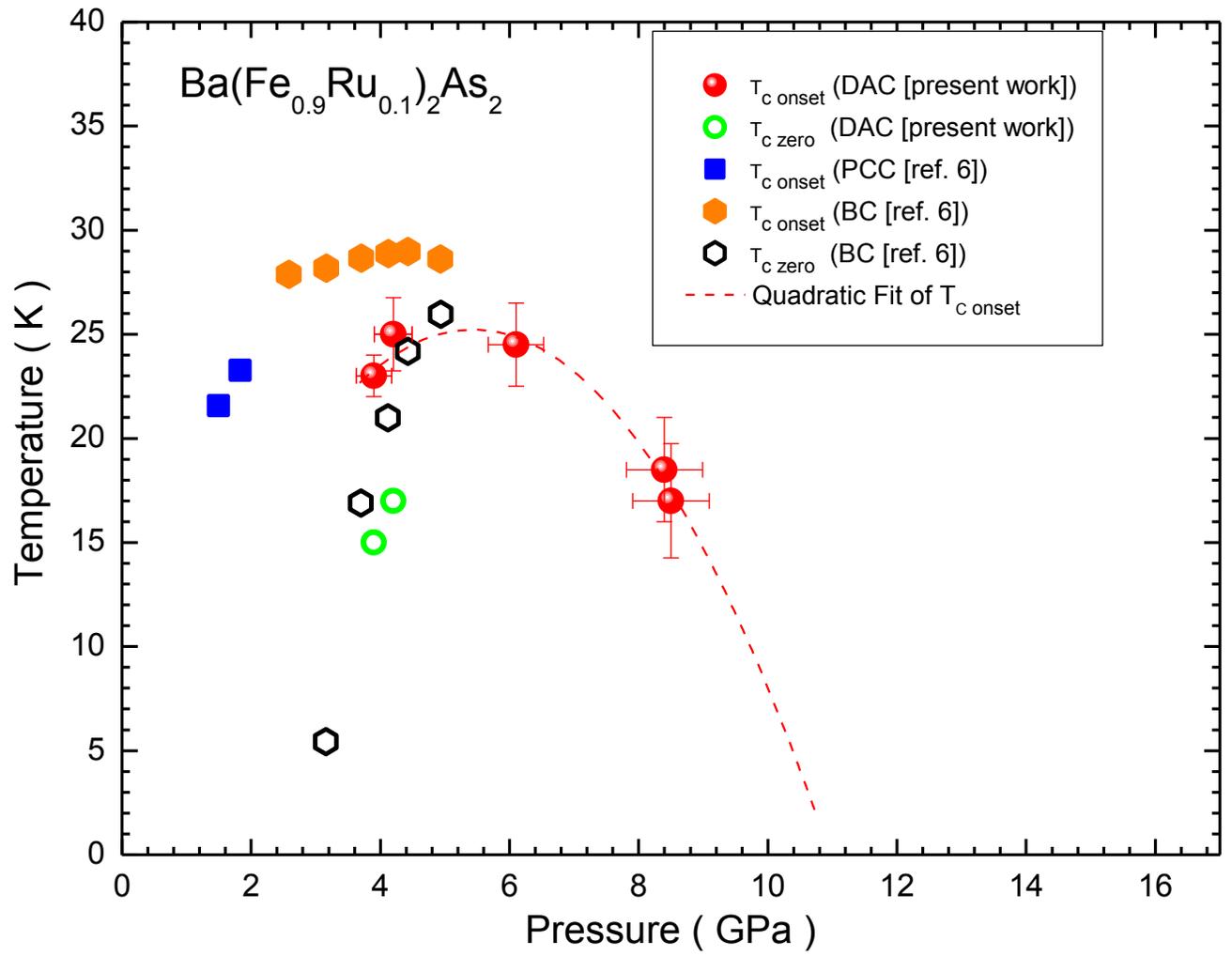

Figure 4